\documentclass[aps,pra,preprint,groupedaddress,showpacs]{revtex4}
\usepackage{graphicx} 
\usepackage{dcolumn}  
\begin{document}
\preprint{Version 3.0}

\title{Two-loop corrections to the decay rate of parapositronium}

\author{Gregory S. Adkins}
\email[]{gadkins@fandm.edu}
\affiliation{Franklin \& Marshall College, Lancaster, Pennsylvania 17604}

\author{Richard N. Fell}
\email[]{fell@brandeis.edu}
\affiliation{Brandeis University, Waltham, Massachusetts 01742}

\author{Nathan M. McGovern}
\affiliation{Franklin \& Marshall College, Lancaster, Pennsylvania 17604}

\author{J. Sapirstein}
\email[]{jsapirst@nd.edu}
\affiliation{University of Notre Dame, Notre Dame, Indiana 46556}

\date{\today}

\begin{abstract}
Order $\alpha^2$ corrections to the decay rate of parapositronium are calculated. A QED
scattering calculation of the amplitude for electron-positron annihilation into two photons at
threshold is combined with the technique of effective field theory to determine an NRQED
Hamiltonian,  which is then used in a bound state calculation to determine the decay rate.
Our result for the two-loop correction is $5.1243(33)$ in units of $(\alpha/\pi)^2$ times the
lowest order rate.  This is consistent with but more precise than the result $5.1(3)$ of a
previous calculation.
\end{abstract}


\maketitle

\section{Introduction}

Effective field theories have proved to be a powerful tool in a variety of applications,
ranging over areas as diverse as nuclear physics \cite{effnuc}, lattice QCD \cite{efflat}, 
heavy flavor physics \cite{effhq}, and Bose-Einstein condensation \cite{effbe}.  One of the 
first  applications of the technique \cite{Caswell-Lepage} was to the bound state problem in
Quantum  Electrodynamics (QED).  As the approach incorporates QED effects as perturbations to
a nonrelativistic  Schr\"{o}dinger problem, it is known as nonrelativistic QED (NRQED). 

Effective field theories in general provide a way of treating physics that involves multiple
scales.  Specifically in the case of atomic physics, NRQED deals with the fact that three
scales--the rest mass of the electron, $m$; the average three-momentum of an electron, $m
\alpha$; and the electron binding energy, $m \alpha^2$--all play significant roles in
radiative and recoil corrections.  While the Bethe-Salpeter equation \cite{Bethe-Salpeter} or
three-dimensional variants of it \cite{3d1,3d2,3d3} provide a consistent framework to carry out
calculations of atomic properties, the implementation is sufficiently complicated so that while 
first-order calculations were carried out in the early 1950's for a number of systems, the next
order calculations were not completely evaluated for over 40 years.

This situation has radically changed since the introduction of effective field theory
techniques for QED bound state calculations.  A set of calculations using various
implementations of NRQED that complete the evaluation of order $\alpha^4$ Rydbergs energy shifts
have  been
carried out over the last few years: we note progress in helium \cite{heliumef},  positronium
\cite{positroniumef}, and muonium
\cite{muoniumef}. In addition to this work on energy levels, the order $\alpha^2$ corrections
to the decay rate of orthopositronium \cite{Annals} and parapositronium \cite{CMY} have also
been treated using NRQED.

In this paper we will be concerned with the latter decay rate, $\Gamma_{p-{\rm Ps}}$.
The dominant decay mode of $p$-Ps is into two photons, and the associated decay rate 
was found by Wheeler \cite{Wheeler} and Pirenne \cite{Pirenne} to be
\begin{equation}
\Gamma_{p-{\rm Ps}}^{(0)}(2\gamma) = {1 \over 2} {m c^2 \over \hbar} \alpha^5 = 
2 \pi c\, {\rm R}_{\infty} \alpha^3 = 8032.5028(1) \mu s^{-1}.
\end{equation}
The numerical value is determined using the 1998 CODATA adjustment
of constants \cite{CODATA}, with the 12 ppb uncertainty dominated
by the uncertainty in the fine structure constant. In the following we will
refer to this lowest order result simply as $\Gamma_0$.

The lowest order rate differs by 0.52 percent from the most accurate measurement \cite{Gidley},
which determines
\begin{equation}
\Gamma_{p-{\rm Ps}}^{exp} = 7990.9(1.7) \mu {\rm s}^{-1}.
\end{equation}
The bulk of the difference is accounted for by the one-loop corrections to the decay
rate, calculated by Harris and Brown \cite{Harris-Brown}, which change the theoretical
prediction by 0.59 percent to
\begin{equation}
\Gamma({\rm 1loop}) = \Gamma_0 \Bigl \{ 1 + { \alpha \over \pi} \Bigl ( {\pi^2 \over 4} - 5
\Bigr ) \Bigr \} =  7985.249 \mu {\rm s}^{-1}.
\end{equation}
The residual 0.07 percent discrepancy corresponds to
a three standard deviation difference between experiment and theory at the one
loop order, and is of a size compatible with order $\alpha^2$ corrections.

Corrections of this order arise both from two-loop corrections--the subject of this    
paper--and from the four photon decay of $p$-Ps. The experiment \cite{Gidley} measures the
total decay rate of $p$-Ps, which in QED includes decays into any even number of photons. We
note that exotic interactions could allow decay into an odd number of photons, and experiments
looking for this kind of decay have put limits on the branching ratio arising from such
interactions, specifically $2.8 \times 10^{-6}$ for $p$-Ps $\rightarrow 3 \gamma$
\cite{Mills} and $2.7 \times 10^{-7}$ for $p$-Ps $\rightarrow 5 \gamma$ \cite{Vetter}. The
only effect that contributes at a non-negligible level is four-photon decay, which was first
calculated in Ref. \cite{Billoire}. The highest accuracy determination of the rate, along with
a calculation of first-order radiative corrections to it, is given in Ref. \cite{Pfahl}, where
references to other calculations can be found. The result is
\begin{equation}
\Gamma_{p-{\rm Ps}}(4\gamma) = 0.274 290(8) \Bigl ({\alpha \over \pi} \Bigr )^2 \Gamma_0 
\Bigl \{ 1 - 14.5(6) \Bigl ( {\alpha \over \pi} \Bigr ) \Bigr \}.
\end{equation}
While this effect is well under the present level of experimental precision, we will
include it in our final prediction.

The remaining contribution from two and higher loops can be parameterized via
\begin{equation}
\Gamma_{p-{\rm Ps}}({\rm 2_+loop}) = \Gamma_0 \Bigl \{ -2 \alpha^2 \ln \alpha + B_{2 \gamma}
\Bigl ( {\alpha \over \pi} \Bigr )^2 - {{3 \alpha^3} \over {2 \pi}} \ln^2 \alpha + C
{{\alpha^3} \over \pi} \ln \alpha + D \Bigl ( {\alpha \over \pi} \Bigr )^3 \Bigr \}.
\end{equation}
The $O(\alpha^2)$ logarithmic term was calculated in Ref. \cite{Khriplovich}, and the
leading $O(\alpha^3)$ logarithm in Ref. \cite{Karshenboim}.  The coefficient of the subleading
$O(\alpha^3)$ logarithmic term
\begin{equation}
C = -{{\pi^2} \over 2} + 10 \ln 2 + {{533} \over {90}} = 7.9189
\end{equation}
is also now well-established, with three different groups \cite{Lepage-Hill,Penin,Melnikov} in
agreement. The terms of order $\alpha^3 \Gamma_0$ are well below the experimental
accuracy, but will be included in our final tally. The leading logarithmic term removes almost
all the residual 0.07 percent discrepancy, so as long as the constant $B_{2 \gamma}$ is not
too large, theory and experiment are in agreement. The
experimental error corresponds to a value of 39 for $B_{2 \gamma}$. Nevertheless, a
direct calculation of this constant is desirable, both because it is possible that the
constant is large, as is frequently the case in QED bound state  calculations, and for
comparison with future experiments of higher precision. As mentioned above, this calculation
has been recently carried out using effective field theory techniques by Czarnecki, Melnikov,
and Yelkhovsky
\cite{CMY}, referred to here as CMY.  After  correction of one part of their calculation
\cite{AFS1}, their result is
\begin{equation}
B_{2 \gamma}=5.1(3).
\end{equation}
This gives for the total theory, including the four-photon decay, the prediction
\begin{equation}
\Gamma^{th}_{p-{\rm PS}}({\rm CMY}) = 7989.616(13) \mu s^{-1},
\end{equation}
which is in good agreement with experiment. The principal result of this paper is the
confirmation of the previous calculation, but with higher precision,
\begin{equation}
B_{2 \gamma}=5.1243(33),
\label{mainresult}
\end{equation}
which leads to our main result,
\begin{equation}
\Gamma^{th}_{p-{\rm Ps}} = 7989.6178(2) \mu s^{-1}.
\end{equation}

The plan of this paper is the following. In section II we explain our implementation of NRQED,
which differs significantly from that of CMY. In section III we carry out a QED calculation of
the scattering amplitude at threshold to two loops.  In the concluding section we use this
amplitude in a bound state calculation to determine $B_{2 \gamma}$, and we discuss the related
$o$-Ps decay rate calculation.

\section{Nonrelativistic Quantum Electrodynamics}

The basic idea of NRQED is to take advantage of the fact that many of the complexities
of QED are associated with the scale of the electron Compton wavelength, which in an atomic
bound state is effectively a point interaction, so that in a bound state calculation one
can account for most QED effects by introducing a delta function potential along with the
usual relativistic perturbations. However, ultraviolet divergences are present when these
interactions are treated in higher order, and the regularization of these divergences can be done 
in different ways. In addition, when carrying out scattering calculations, the infinite range
of the Coulomb interaction leads to infrared divergences that can also be regularized in
different ways. For these reasons, the details of an NRQED calculation can be quite different 
when done by different groups. We consider this to be an advantage, as agreement between
different methods, such as will be found here, lends support to the reliability of these
complex calculations. 

CMY handled both ultraviolet and infrared divergences with dimensional regularization.  In the
present work, we only use dimensional regularization to handle ultraviolet divergences in the
QED scattering calculation of section III.  We regulate infrared divergences by introducing a
photon mass $\tilde{\lambda} = m \lambda$, and we cut off ultraviolet divergences in the
NRQED calculation with a maximum loop three-momentum $\Lambda$.

A detailed description of NRQED applied to the decay rate of orthopositronium
can be found in Ref. \cite{Annals}. A great deal of the analysis given there applies
equally well to parapositronium, but for completeness we give here a complete, but
somewhat abbreviated, discussion, referring the reader interested in more details to
Ref. \cite{Annals}.

Because we regulate infrared infinities with a photon mass, it is necessary to set up
a consistent set of interactions that incorporate this effect. These interactions are
most conveniently expressed in the center-of-mass frame in momentum space, with an incoming
electron momentum $\vec k$ and an outgoing electron momentum $\vec l$. It is also
useful to define the frequently occurring denominators
\begin{equation}
D_{\lambda}(\vec k) = \vec k^2 + \tilde{\lambda}^2
\end{equation}
and
\begin{equation}
D_k = \vec k^2 + \gamma^2,
\end{equation}
where $\gamma = m \alpha /2$.  The NRQED interactions are depicted in Fig.~1.

\begin{figure}
\includegraphics{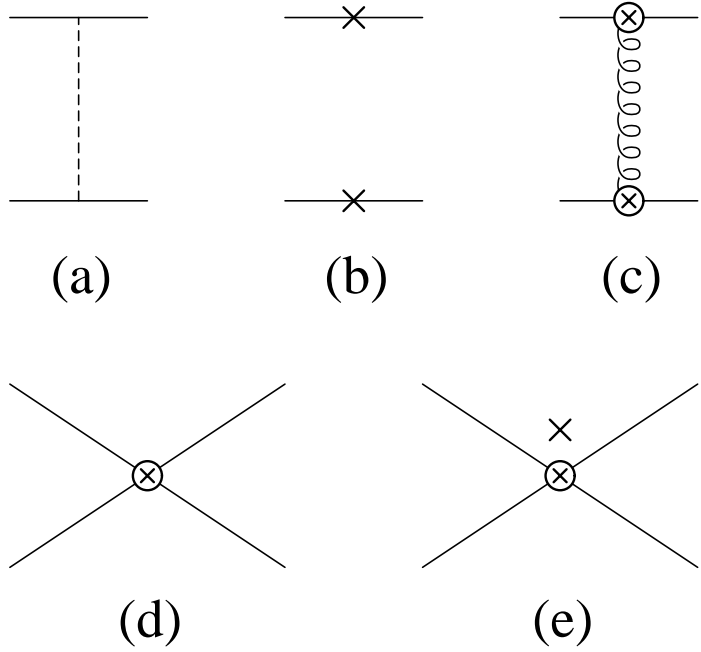}
\caption{\label{fig1} The NRQED instantaneous potentials: (a) Coulomb, (b) relativistic mass
increase, (c) Breit-Fermi, (d) four-fermion contact, (e) four-fermion derivative.}
\end{figure}

At lowest order, the NRQED Hamiltonian consists of the usual nonrelativistic kinetic energy 
together with a modified Coulomb potential,
\begin{equation}
V_C( \vec k, \vec l \,) = - {4 \pi \alpha \over D_{\lambda} (\vec k - \vec l \,)}.
\end{equation}
The usual relativistic perturbations responsible for the fine structure of hydrogen 
are present, including the relativistic mass increase (RMI),
\begin{equation}
V_{RMI}(\vec k, \vec l \,) = - (2 \pi)^3 \delta(\vec k - \vec l \,) {\vec k^4 \over 4 m^3}
\end{equation}
and a term, which includes both relativistic corrections to the Coulomb potential and the
exchange of a transverse photon, denoted $V_{BF}$. The full form of this
``Breit-Fermi'' interaction when a photon mass is present is
\begin{eqnarray}
V_{BF}(\vec k, \vec l \,) & = & - {4 \pi \alpha \over m^2} \Bigl [ { |\vec k \times \vec l
\,|^2 + {1 \over 4}  \tilde{\lambda}^2
|\vec k + \vec l \,|^2 \over D_{\lambda}^2(\vec k - \vec l \,)} -
{(\vec k - \vec l \,) \times \vec S_{-} \cdot (\vec k - \vec l \,)\times \vec S_{+} \over
D_{\lambda}(\vec k - \vec l \,)} \nonumber \\
& + & {3 \over 2} i { (\vec k \times \vec l \,) \cdot (\vec S_{-} + \vec S_{+}) \over
D_{\lambda}(\vec k - \vec l \,)} - {1 \over 4} {|\vec k - \vec l \,|^2 \over 
D_{\lambda}(\vec k - \vec l \,)} \Bigr ].
\end{eqnarray}
It includes the Darwin term and the spin-orbit interaction, though we note the latter
interaction does not contribute for S-states. 

The next interaction we need for NRQED is the four-point interaction describing the
decay process,
\begin{equation}
V_4(\vec k, \vec l \,) = V_4^0 \Bigl \{ 1 + { \alpha \over \pi} e_1 + \Bigl ({\alpha \over \pi}
\Bigr )^2 e_2 \Bigr \}
\end{equation}
where
\begin{equation}
V_4^0 = - { 2 \pi \alpha^2 \over m^2} i.
\end{equation}
The constants $e_1$ and $e_2$ renormalize the interaction and will be determined from a
matching calculation below. This interaction corresponds to a delta function in coordinate
space and gives rise to an  energy in which $V_4$ is multiplied by the square of the wave
function at the origin, $m^3 \alpha^3 /8 \pi$. The leading term is chosen so that the decay
rate, given by $\Gamma = - 2 Im (E)$, reproduces the known lowest order result.

An important difference of the $p$-Ps calculation from the $o$-Ps calculation is the fact that the
latter has a real contribution to $V_4$ arising from one-photon annihilation, a channel not
available in the present case. The last interaction needed for the calculation accounts for the
fact that
the annihilation is not exactly pointlike, which leads to a derivative term,
\begin{equation}
V_{4der}(\vec k, \vec l \,) =  {4 \pi \alpha^2 \over 3 m^4} i ( \vec k^2 + \vec l \,^2).
\end{equation}
We note the relationship
\begin{equation}
V_{4der} (\vec k, \vec l \,) = - {2 \over 3} V_{4}^0 { \vec k^2 + \vec l \,^2 \over m^2}
\end{equation}
happens to be identical to the behavior of the interaction for $o$-Ps arising from
one-photon exchange,
which contributes to the hfs of positronium; this fact will be used below.  We now describe how
these interactions are used in a bound state NRQED  calculation.

\subsection{Bound state calculation}

\begin{figure}
\includegraphics{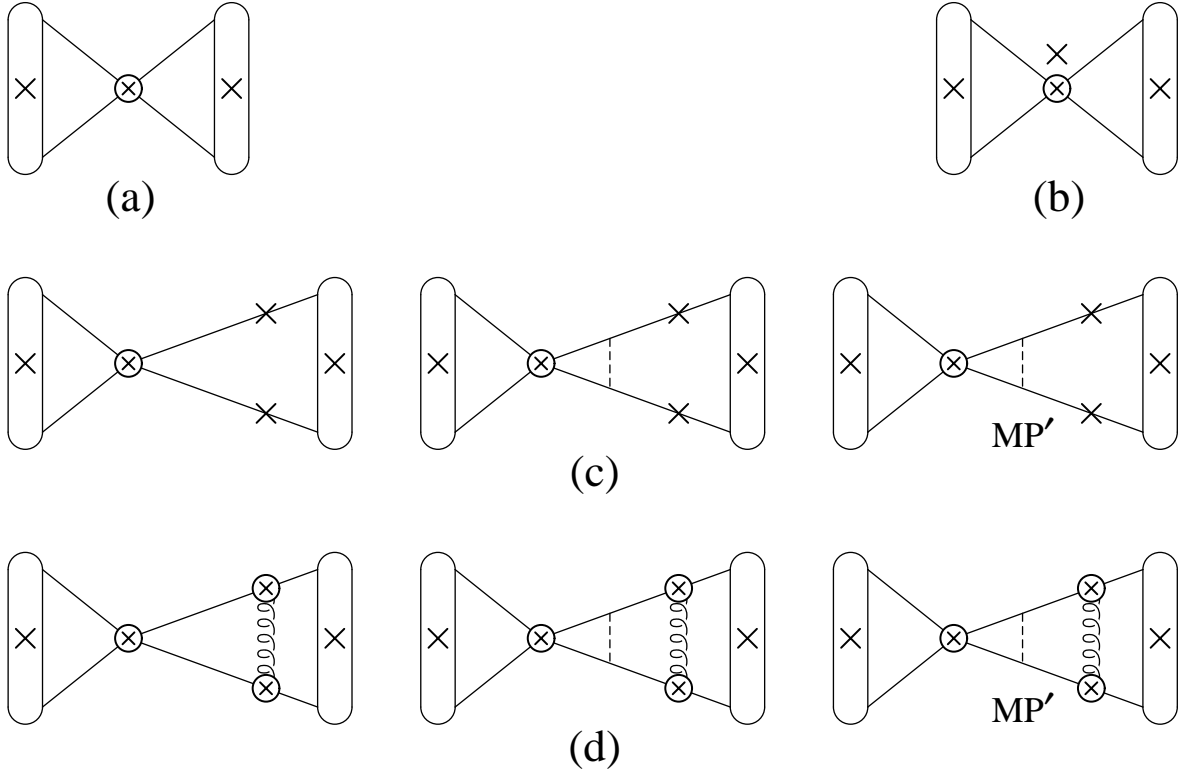}
\caption{\label{fig2}  First and second order bound state perturbation theory contributions to
the energy shift.  They are the first order (a) contact and (b) derivative contributions, and
the second order (c) relativistic mass increase and (d) Breit-Fermi contributions. 
Contributions (c) and (d) are shown separated into their zero-, one-, and 
many-potential parts.}
\end{figure}

We now apply standard Rayleigh-Schr\"{o}dinger perturbation theory to second order.  (The
corresponding diagrams are shown in Fig.~2.)  We are interested only in imaginary contributions
to the energy, so one factor of $V_4$ or $V_{4der}$ must be present. The expression for first
order perturbation theory in momentum  space is
\begin{equation}
E^{(1)}_V = \int {d^3 p_2 \over (2\pi)^3} {d^3 p_1 \over (2\pi)^3} \phi^*(\vec p_2) V(\vec p_2,
\vec p_1) \phi(\vec p_1),
\end{equation}
where the wave function is
\begin{equation}
\phi(\vec p \,) =  \Bigl ( {\gamma^3 \over \pi} \Bigr )^{1/2} { 8 \pi \gamma \over ({\vec p
\,}^2 + \gamma^2)^2}.
\end{equation}
This gives the contributions
\begin{equation}
E^{(1)}_{V_4} = - {i \over 4} m \alpha^5 \Bigl \{ 1 + {\alpha \over \pi} e_1 + \Bigl ({\alpha
\over \pi} \Bigr )^2 e_2 \Bigr \}
\end{equation}
from Fig.~2a and 
\begin{equation}
E^{(1)}_{V_{4der}} = { 2 i \Lambda \alpha^6 \over 3 \pi} - {1 \over 4} i m \alpha^7
\end{equation}
from Fig.~2b, where $\Lambda$ is the maximum value allowed for the magnitude of the
three-momentum $|\vec p_2|$ or $|\vec p_1|$.

In second order we need to evaluate expressions of the form
\begin{equation}
E^{(2)}_{ij} = \sum_{n \neq 0} { <0 | V_i |n > < n | V_j| 0> \over E_0 - E_n },
\end{equation}
which can be written in terms of a reduced Coulomb Green's function,
\begin{equation}
E^{(2)}_{ij} = {1 \over (2 \pi)^{12}} \int d^3 p_2 \, d^3 k \, d^3 l \,
d^3 p_1 \, \phi^*(\vec p_2) V_i(\vec p_2, \vec k)
G_R(\vec k, \vec l \,) V_j(\vec l, \vec p_1) \phi(\vec p_1).
\end{equation}
The reduced Green's function in momentum space conveniently breaks into
three parts, in which the electron either propagates freely, interacts
once with a Coulomb potential, or interacts more than once.  It is given by
\begin{equation}
G(\vec k , \vec l \,) = - (2 \pi)^3 \delta^3( \vec k - \vec l \,) { m \over D_k}
- { m \over D_k} {4 \pi \alpha \over | \vec k - \vec l \,|^2}
{ m \over D_l}  - R( \vec k , \vec l \,)
\end{equation}
where
\begin{equation}
R(\vec k , \vec l \,) = { 32 \pi m \gamma^3 \over D_k^2 D_l^2}
\Bigl [ { 5 \over 2} - {4 \gamma^2 \over D_k} - {4 \gamma^2 \over D_l} + {1 \over 2} \ln A
+ {2A-1 \over \sqrt{4A-1}} {\rm tan}^{-1}\sqrt{4A-1} \Bigr ],
\end{equation}
with
\begin{equation}
A= { D_k D_l \over 4 \gamma^2 | \vec k - \vec l \,|^2}.
\end{equation}

When the interaction is a relativistic mass increase (see Fig.~2c), second order perturbation
theory gives
\begin{equation}
E^{(2)}_{V_4,V_{RMI}}+E^{(2)}_{V_{RMI},V_4} = -i {\Lambda \alpha^6 \over 4 \pi}
- i { m \alpha^7 \over 8} \ln \Bigl ( {\Lambda \over 2 \gamma} \Bigr )
- i { m \alpha^7 \over 32},
\end{equation}
and when it is a Breit-Fermi interaction (see Fig.~2d), the contribution is
\begin{equation}
E^{(2)}_{V_4,V_{BF}} + E^{(2)}_{V_{BF},V_4} = -i {\Lambda \alpha^6 \over 4 \pi}
- i { 3 m \alpha^7 \over 8} \ln \Bigl ( {\Lambda \over 2 \gamma} \Bigr )
- i { 7 m \alpha^7 \over 16},
\end{equation}
where contributions of order $m \alpha^8$ and higher are dropped.
A table of integrals useful in arriving at these results can be found in Ref. \cite{Annals}.

The final result for the NRQED bound state calculation is then
\begin{equation}
\Gamma_{\rm NRQED} = \Bigl \{ 1 + { \alpha \over \pi} \Bigl [ e_1 - {2 \Lambda \over 3
m} \Bigr ]+ \Bigl ( {\alpha \over \pi} \Bigr )^2 \Bigl [ e_2 
+ {23 \pi^2 \over 8} + 2 \pi^2 \ln \Bigl ( {\Lambda \over 2 \gamma} \Bigr ) \Bigr ] \Bigr \}
\Gamma_0,
\label{BSNRQED}
\end{equation}
which has been converted to a decay rate in the usual manner.

\subsection{NRQED scattering calculation}

The next step in the analysis is to consider the imaginary part of the amplitude 
in NRQED for Bhabha scattering at threshold. In lowest order this is simply $V_4^0$, 
with no contribution from $V_{4der}$ because of the presence of the momentum factors.
Including the renormalization terms, we denote the overall effect of $V_4$ as
\begin{equation}
M^{(0)}_{NRQED} = V_4^0 \Bigl \{ 1 +  {\alpha \over \pi} e_1 + \Bigl ( {\alpha \over \pi}
\Bigr )^2 e_2 \Bigr \}.
\end{equation}
In first order, a simple calculation gives the contributions
\begin{equation}
M^{(1)}_{NRQED}(V_C) = { \alpha \over \pi} V_4^0 \Bigl ( {{2 \pi} \over \lambda} \Bigr
) \Bigl \{ 1 + {\alpha \over \pi} e_1 \Bigr \},
\end{equation}
\begin{equation}
M^{(1)}_{NRQED}(V_{RMI}) = {\alpha \over \pi} V_4^0 \Bigl ({\Lambda \over m} - {\pi
\lambda \over 2} \Bigr ),
\end{equation}
\begin{equation}
M^{(1)}_{NRQED}(V_{BF}) = { \alpha \over \pi} V_4^0 \Bigl ({\Lambda \over m} - {\pi
\lambda \over 4} \Bigr ),
\end{equation}
and
\begin{equation}
M^{(1)}_{NRQED}(V_{4der}) = { \alpha \over \pi} V_4^0 \Bigl (-{8 \Lambda \over 3 m} + 
{4 \pi \lambda \over 3} \Bigr ),
\end{equation}
shown in Fig.~3 a-d, which sum to
\begin{equation}
M^{(1)}_{NRQED} = { \alpha \over \pi} V_4^0 \Bigl ( { 2 \pi \over \lambda} \Bigl \{ 1 +
{\alpha \over \pi} e_1 \Bigr \} -{2 \Lambda \over 3 m} +  {7 \pi \lambda \over 12} \Bigr ).
\label{nrqed1}
\end{equation}

\begin{figure}
\includegraphics{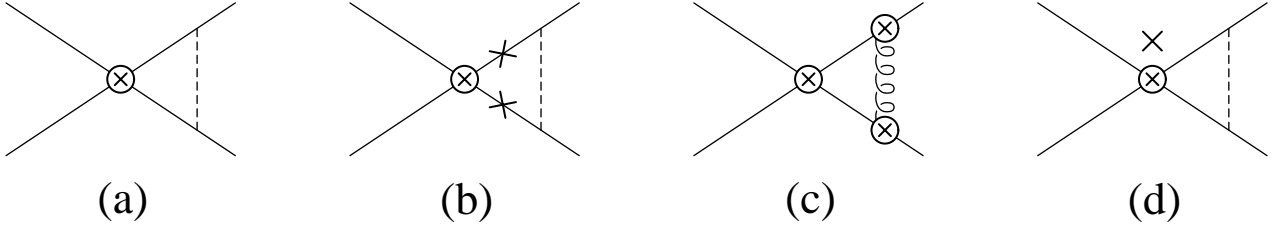}
\caption{\label{fig3} The one-loop NRQED $e^-e^+ \rightarrow e^-e^+$ scattering graphs.}
\end{figure}

In the next section we will require this amplitude equal the one-loop QED amplitude to determine
$e_1$. To determine $e_2$ a two-loop calculation is required. The calculation is almost identical
to the scattering calculation carried out for hyperfine splitting in section 3 of
Ref. \cite{Annals}, where twelve diagrams (labeled a-l) were analyzed. The last two of them,
$k$ and $l$,  involved a real contribution to $V_4$ not present for $p$-Ps, so they are
dropped. The other ten (see Fig.~4) are all  proportional to the lowest order amplitude and
can be taken over directly to $p$-Ps by simply using  the lowest order amplitude for $p$-Ps
instead of hfs. The only other change needed is the modification of the three diagrams which
involve the $V_{BF}$ interaction, which is different for the two states. However, the
difference is simply
\begin{equation}
\delta V_{BF} = - { 8 \pi \alpha \over 3 m^2} { |\vec k - \vec l \,|^2 \over D_{\lambda}(\vec k
-\vec l \,)},
\end{equation}
which leads to the modifications
\begin{equation}
\delta f = \Bigl ( {\alpha \over \pi} \Bigr )^2 V_4^0 \Bigl ({ 8 \pi \Lambda \over 3
\tilde{\lambda}} -  {4 \pi^2 \ln 2 \over 3} \Bigr ),
\end{equation}
\begin{equation}
\delta g = \Bigl ( {\alpha \over \pi} \Bigr )^2 V_4^0 { 4 \pi ^2 \over 3} \ln \Bigl ( {\Lambda
\over 2 \tilde{\lambda}} \Bigr ),
\end{equation}
and
\begin{equation}
\delta h = \Bigl ( {\alpha \over \pi} \Bigr )^2 V_4^0 \Bigl ({ 8 \pi \Lambda \over 3
\tilde{\lambda}} - {4
\pi^2 \over 3} \Bigr ).
\end{equation}

\begin{figure}
\includegraphics{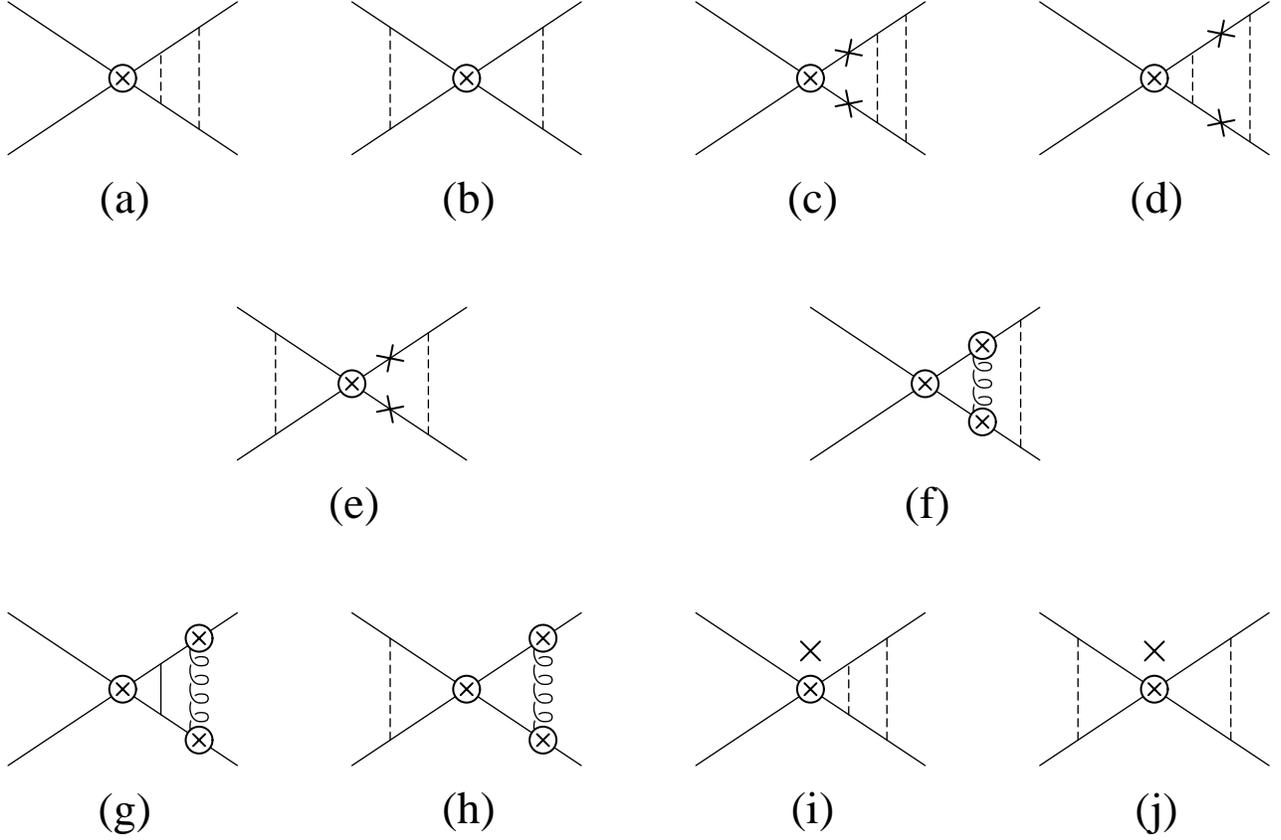}
\caption{\label{fig4} The two-loop NRQED $e^-e^+ \rightarrow e^-e^+$ scattering graphs.}
\end{figure}

Finally, graphs $i$ and $j$ are derivative terms, but as mentioned earlier, the ratio of the
derivative term to the lowest order matrix element is the same for hfs as it is for $p$-Ps,
so the results from Ref. \cite{Annals} can be taken over directly. The final result for the
two-loop contribution to threshold scattering is
\begin{equation}
M^{(2)}_{NRQED} = \Bigl ({\alpha \over \pi} \Bigr )^2 V_4^0 \Bigl ( {\pi^2 (2 \ln 2 + 1) \over
\lambda^2} -{4 \pi \Lambda \over 3 \tilde{\lambda}} + 
2 \pi^2 \ln \Bigl ( { \Lambda \over 2 \tilde{\lambda}} \Bigr ) + { 4 \pi^2 \over 3} \Bigr ),
\end{equation}
and the NRQED amplitude that is to be compared to the QED scattering calculation is 
\begin{equation}
M_{NRQED} = M^{(0)}_{NRQED} + M^{(1)}_{NRQED} + M^{(2)}_{NRQED}.
\label{nrqed2}
\end{equation}

\section{QED scattering calculation}

We are interested in the imaginary part of the scattering amplitude for an
electron and positron to annihilate into two photons, and the two photons then
to create an electron and positron. While this amplitude at threshold is simply
a constant, in order to determine the derivative term in NRQED we also need the behavior 
slightly above threshold. If we assign the momentum $\vec k$ to the incoming electron
and $\vec l$ to the outgoing electron, with the positrons having opposite momentum,
a straightforward calculation leads to the above-threshold matrix element
\begin{equation}
M^{(0)}_{QED} (\vec k, \vec l \,) = V_4^0 \Bigl ( 1 - {2 \over 3} { \vec k^2 + \vec l \,^2
\over m^2} \Bigr ).
\end{equation}
This requires that NRQED have the forms for $V_4$ and $V_{4der}$ used
in the previous section. At threshold we have simply
\begin{equation}
M^{(0)}_{QED}  = V_4^0.
\end{equation}
We will follow the convention of listing results for a given $n$-loop contribution to the
scattering amplitude
\begin{equation}
M_{QED}^{(n)} = \Bigl ( {\alpha \over \pi} \Bigr )^n V_4^0 I^{(n)}
\end{equation}
by giving $I$ instead of the full amplitude.  So for the zero-loop contribution we have
\begin{equation}
I^{(0)} = 1.
\end{equation}

Continuing to the one-loop calculation, there are three diagrams that must be considered,
which we call the self-energy (SE), vertex (V), and ladder (L). They are
shown in Fig.~5.  The self-energy and vertex results are
\begin{figure}
\includegraphics{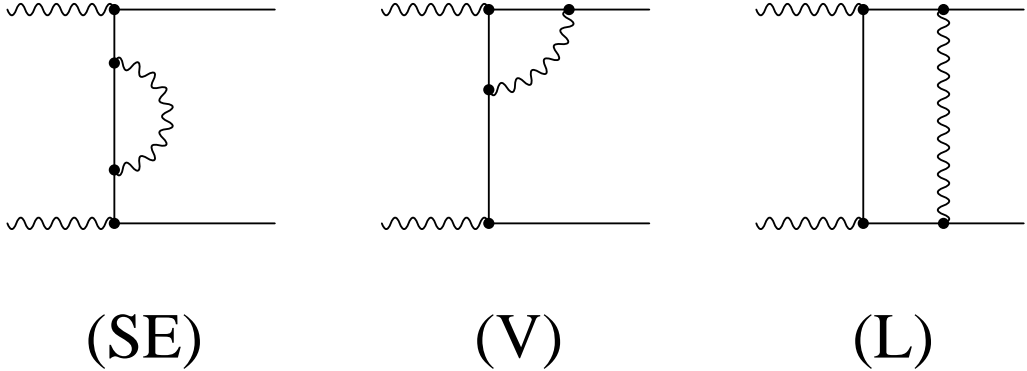}
\caption{\label{fig5} One-loop QED graphs contributing to the parapositronium decay rate.  They
are (SE) the self-energy graph, (V) the vertex graph, and (L) the ladder graph.}
\end{figure}

\begin{equation}
I_{SE} = 2 \ln \lambda + 4 \ln 2 + 1 + O(\lambda),
\label{oneloopse}
\end{equation}
and
\begin{equation}
I_V = -4 \ln \lambda + {{\pi^2} \over 4} - 4 \ln 2 - 4 + O(\lambda).
\label{oneloopv}
\end{equation}
The infrared logarithms present in $I_{SE}$ and $I_V$ are artifacts of the renormalization
process.  The ladder diagram has an infrared singularity associated with binding that we denote
as
$I_B$, defined by
\begin{equation}
I_B = {\pi \over \lambda} + \ln \lambda -1 - {\pi \lambda \over 8} + O(\lambda^2).
\end{equation}
The ladder diagram turns out to be proportional to $I_B$, which will also prove
a useful factor in the two-loop calculation.  We find
\begin{equation}
I_{L} = 2 I_B + O(\lambda).
\label{oneloopladder}
\end{equation}
A simple summation of the one-loop contributions gives the overall result
\begin{equation}
I^{(1)} = {2 \pi \over \lambda} +  A(\lambda),
\label{gamma1qed}
\end{equation}
with 
\begin{equation}
A(\lambda) = {\pi^2 \over 4} -5 + E \lambda.
\end{equation}
Here $E \lambda$ represents the uncalculated order $\lambda$ contribution.  The value of $E$
does not affect the final result.

The two-loop calculation involves eight sets of graphs. In this part of the calculation
ultraviolet divergences are regulated by working in $n = 4 - 2 \epsilon$ dimensions.  When
doing so a common factor of $[(4 \pi \mu^2/m^2) e^{-\gamma_E}]^{2 \epsilon}$ is always present
(where $\mu$ is a mass scale introduced in the process of dimensional regularization, and
$\gamma_E$ is the Euler gamma constant), but we leave it unwritten. It can be taken to unity
after renormalization.

We begin the discussion with two-loop corrections to a photon vertex, shown in Fig.~6a. In
diagrams a2 and a5 the self-energy subdiagram is understood to be accompanied by a self-mass
counterterm.  This convention differs from CMY, but we have been able to show agreement with
their results if the counterterm is included. The unrenormalized results are given in Table~I. 
The total class a unrenormalized result is
\begin{figure}
\includegraphics{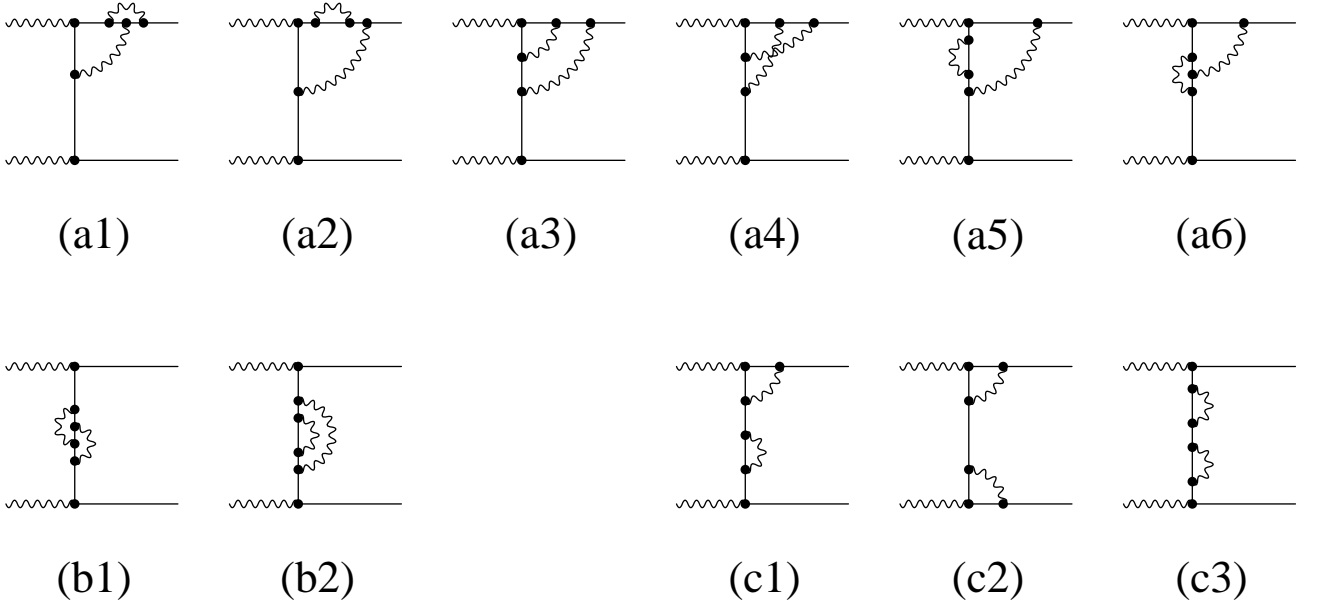}
\caption{\label{fig6} Contribution to the p-Ps decay rate from groups a (the two-loop vertex
corrections), b (the two-loop self-energy corrections), and c (products of one-loop vertex
and self-energy parts).}
\end{figure}
\begin{equation}
I_a' = {1 \over {8 \epsilon^2}} + {1 \over \epsilon} \Bigl ( {{\pi^2} \over {16}} - \ln 2 - 
{3 \over {16}} \Bigr ) + 2.8841255(52).
\end{equation}
This is renormalized according to
\begin{equation}
I_a = I_a' - \ell_1 I_V' - 4(\tilde{\ell}_2-\ell_1^2) I^{(0)},
\end{equation}
where
\begin{equation}
I_V' = {1 \over \epsilon} + \Bigl ({{\pi^2} \over 4} - 4
\ln 2 \Bigr ) + \epsilon \Bigl ( {7 \over 2} \zeta(3) + {{\pi^2} \over 4} + 4 \ln^2 2 - 8 \ln
2 \Bigr ) + O(\epsilon^2)
\end{equation}
is the unrenormalized one-loop vertex correction (see Fig.~5); $L_1=(\alpha / \pi) \ell_1$,
where
\begin{equation}
\ell_1 = {1 \over {4 \epsilon}} + \Bigl ( \ln \lambda + 1 \Bigr ) +
\epsilon \Bigl ( -\ln^2 \lambda + {{\pi^2} \over {48}} + 2 \Bigr ) + O(\epsilon^2),
\end{equation}
is the one-loop vertex renormalization constant; and $\tilde{L}_2 = (\alpha / \pi)^2
\tilde{\ell}_2$, where
\begin{equation}
\tilde{\ell}_2 = {1 \over {32 \epsilon^2}} + {1 \over \epsilon} \Bigl ( {{\ln \lambda} \over
4} + {{13} \over {64}} \Bigr ) + \Bigl ( {1 \over 4} \ln^2 \lambda + \ln \lambda + {3 \over 2}
\zeta(3) - \pi^2 \ln 2 + {{157 \pi^2} \over {192}} - {{49} \over {128}} \Bigr ) +
O(\epsilon),
\end{equation}
is the two-loop renormalization constant (with the vacuum polarization effect excluded). 
Because we treat vacuum polarization separately, it is important to again emphasize that the
renormalization constant above also does not include the effect of vacuum polarization.  Our
result for the two-loop renormalized class a contribution is
\begin{equation}
I_a = -2 \ln^2 \lambda - \ln \lambda I_V(\lambda) - 1.966447(6),
\end{equation}
where $I_V(\lambda)$ is the one-loop vertex factor of Eq.~(\ref{oneloopv}).

Turning to the two-loop self-energy diagrams, shown in Fig.~6b, we again present results with
self-mass counterterms understood. The unrenormalized results are given in Table~II and are in
agreement, after accounting for the different convention, with CMY.  The total class b
unrenormalized result is
\begin{equation}
I_b' = -{{1} \over {16 \epsilon^2}} + {1 \over \epsilon} \Bigl ( \ln 2 - {5 \over {32}} \Bigr
) - 1.6727038(15).
\end{equation}
The corresponding renormalized expression is
\begin{eqnarray}
I_b &=& I_b' - \ell_1 I_{SE}' + 2 (\tilde{\ell}_2-\ell_1^2) \cr
&=& \ln^2 \lambda - \ln \lambda I_{SE}(\lambda) - 2.0277430(15),
\end{eqnarray}
where $I_{SE}'$ is the unrenormalized one-loop self-energy correction
\begin{equation}
I_{SE}' = -{{1} \over {2 \epsilon}} + \Bigl ( 4 \ln 2 - 1 \Bigr ) + \epsilon \Bigl ( {{7 \pi^2}
\over {24}} - 4 \ln^2 2 + 6 \ln 2 - 2 \Bigr ) + O(\epsilon^2) ,
\end{equation}
and $I_{SE}(\lambda)$ is the renormalized one-loop self-energy factor of Eq.~(\ref{oneloopse}).

The diagrams with two separate one-loop corrections shown in Fig.~6c are simple to evaluate
and can be done entirely analytically. The breakdown of the calculation is given in Table~III
and agrees exactly with CMY.  Interestingly, the total class c contribution is unchanged by
renormalization.  It is
\begin{equation}
I_c = {{\pi^4} \over {128}} + {1 \over 4} \pi^2 \ln 2 - {{\pi^2} \over 8} + \ln^2 2 - \ln 2 +
{1 \over 4} .
\end{equation}

The least complicated contributions are those, shown in Fig.~7d, which have no ultraviolet or
infrared  divergences. The results, given in Table~IV, can be directly compared with the
diagrams called $D_{14}$ and $D_3$ by CMY and are in agreement.  For the class d total we find
\begin{equation}
I_d = -0.87783(7).
\end{equation}
\begin{figure}
\includegraphics{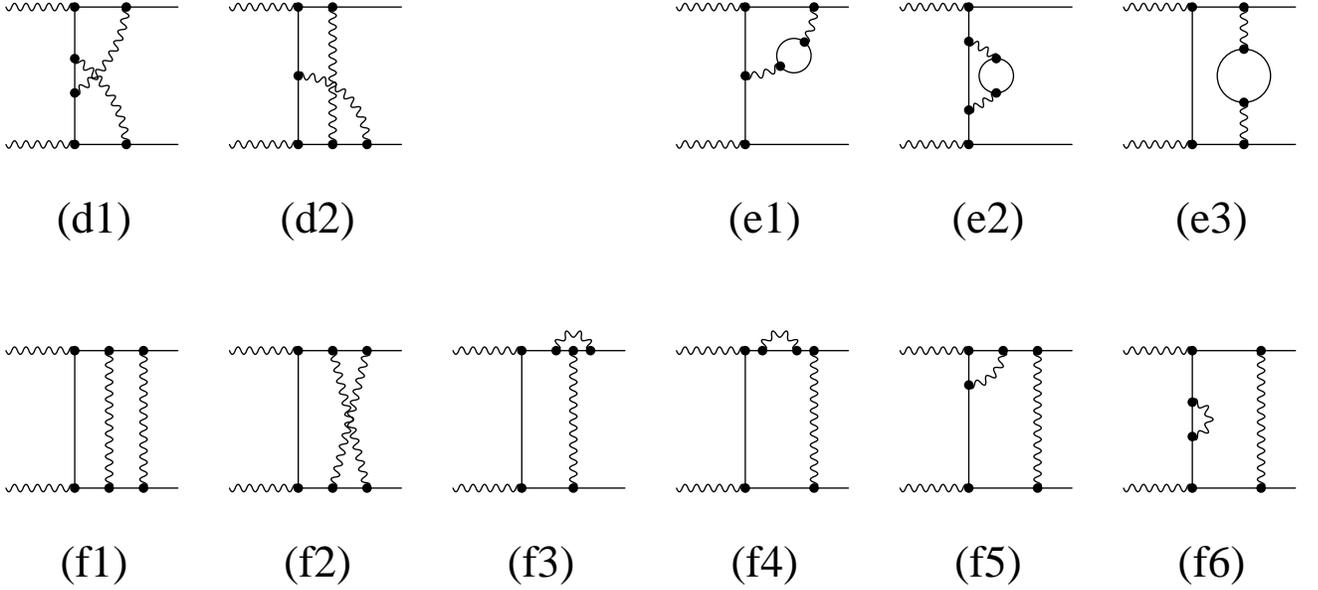}
\caption{\label{fig7} Contribution to the p-Ps decay rate from groups d (infrared and
ultraviolet finite contributions), e (vacuum polarization corrected one-loop graphs), and f
(ladder graphs).}
\end{figure}

The vacuum polarization diagrams of Fig.~7e were evaluated in Refs. \cite{Bur95,AdkinsShif},
with result
\begin{equation}
I_e = 0.4473430(6).
\end{equation}

As with $o$-Ps, by far the most difficult diagrams to treat were those of class f, with at
least one binding photon exchanged.  As before, the most infrared singular contributions are
the ladder and crossed-ladder f1 and f2.  They were treated just as in the $o$-Ps calculation
by subtracting off the most infrared singular part of each (called f1a0 and f2a0), which were
evaluated as a unit.  We found that
\begin{eqnarray}
I_{f1} &=& I_{f1a0} + I_B(\lambda) I_{LS}(\lambda) - 3 \pi^2 \ln \lambda - 55.2277(15), \\
I_{f2} &=& I_{f2a0} + \pi^2 \ln \lambda + 30.8346(7),
\end{eqnarray}
where (see Eq.~(\ref{oneloopladder})) $I_{LS} \equiv I_L-2 I_B=O(\lambda)$.  The leading
binding singularity is contained in
\begin{equation}
I_{f1a0}+I_{f2a0} = 2 I^{a0} = {{2 \pi^2 \ln 2} \over {\lambda^2}} + 2 \Bigl ( I_B-{\pi
\over \lambda} \Bigr ) {\pi \over \lambda} + \ln^2 \lambda - 2 \ln \lambda + 2 A_0,
\end{equation}
where
\begin{equation}
A_0 = -3.2036(13)
\end{equation}
as given in Eqs.~(3.37) and (3.86) of \cite{Annals}.  Diagrams f3 and f4 were evaluated
using Yennie gauge for the vertex or self-energy photon and Feynman gauge for the ladder
photon.  This procedure is consistent since the sum of the two is independent of the gauge of
the vertex or self-energy photon.  An all-Feynman-gauge evaluation of f3 and f4 proved to be
numerically difficult.  The class f results are given in Table~V.  The total class f
contribution is
\begin{equation}
I_f = {{2 \pi^2 \ln 2} \over {\lambda^2}} + {\pi \over \lambda} A(\lambda) + \ln^2 \lambda +
(I_V(\lambda) + I_{SE}(\lambda)-2 \pi^2) \ln \lambda - 35.0357(32).
\end{equation}

Class g is the set of energy shift contributions having one-loop corrections (SE, V, or L)
both before and after the annihilation into two photons.  This contribution is easily
evaluated as the square of half the one-loop correction:
\begin{equation}
I_g = \Bigl [ {1 \over 2} I^{(1)} \Bigr ]^2 = {{\pi^2} \over {\lambda^2}} + {\pi \over
\lambda} A(\lambda) + \Bigl ( {{\pi^2} \over 8} - {5 \over 2} \Bigr )^2 .
\end{equation}

Finally, class h represents the effect of light-by-light scattering on the annihilation
photons.  This contribution was evaluated in Ref.~\cite{AFS1} to be
\begin{equation}
I_h = 1.29392(4).
\end{equation}

The total two-loop contribution is then
\begin{equation}
I^{(2)} = { \pi^2 (2 \ln 2 + 1)  \over \lambda^2}  + {2 \pi \over \lambda} A(\lambda) 
-2 \pi^2 \ln \lambda +  B_2,
\label{gamma2qed}
\end{equation}
with
\begin{equation}
B_2 = -35.2881(33).
\label{QEDres}
\end{equation}
The total QED scattering amplitude at threshold through two-loop order is the sum of the
$M^{(i)}_{QED}$ for $i$ equal to 0, 1, and 2.

Most of the two-loop results can be cross-checked against corresponding results of CMY as
given in Ref.~\cite{CMY}.  This comparison is detailed in Table~VII.  The a graphs can be
compared directly to the corresponding CMY graphs except for a2 and a5, whose CMY counterparts
$D_{11}$ and $D_{12}$ require the inclusion of the one-loop self-mass $\delta m^{(1)}$ as
given in CMY Eq.~(48).  (For ease of notation we take $m \rightarrow 1$ in the CMY expressions
for $\delta m$.)  The CMY counterparts of the b graphs require one- and two-loop self-mass
insertions.  It is necessary to isolate the ``crossed rainbow'' and ``double rainbow''
contributions to $\delta m^{(2)}$, which are
\begin{eqnarray}
\delta m^{(2)}_{CR} &=& -{3 \over {\epsilon^2}} - {5 \over {2 \epsilon}} - 12 \zeta(3) + 8
\pi^2 \ln 2 - {{9 \pi^2} \over 2} + {1 \over 4}, \\
\delta m^{(2)}_{DR} &=& {{15} \over {2 \epsilon^2}} + {{55} \over {4 \epsilon}} + {{\pi^2}
\over 4} + {{197} \over 8},
\end{eqnarray}
as given in Ref.~\cite{Adkins01} but adapted to the conventions of CMY.  The c graphs were done
analytically by both groups, and agree exactly once appropriate self-mass insertions are
made.  The d graphs are directly comparable.  The e (vacuum polarization) graphs were taken
from Ref.~\cite{AdkinsShif} by both groups.  The f graphs, which involve the binding
interaction, were the most difficult to compare.  In fact, the double ladder and crossed
ladder graphs f1 and f2 could not be checked against their CMY counterparts $D_4$ and
$D_1$ because the methods for regulating the binding singularities were completely different. 
We evaluated graphs f3 and f4 with Yennie gauge vertex and self-energy photons, but the sum is
directly comparable with
$D_2+D_5$ of CMY after the appropriate self-mass correction is made to their result.  The
comparison for f5 and f6 is a bit more delicate.  For the sum f5+f6 we found
\begin{equation}
I_{f5}+I_{f6} = (I_V+I_{SE}) I_B + 1.29396(4) ,
\end{equation}
where $I_V$ and $I_{SE}$ are the renormalized vertex and self-energy corrections.  Now CMY
calculate unrenormalized graphs, so we revert to the unrenormalized contribution as well:
$I_V+I_{SE} \rightarrow I_V'+I_{SE}'$, where in CMY notation
\begin{equation}
I_V'+I_{SE}' = 4 \bigl ( S_2+S_3+\delta m^{(1)} B_1 \bigr ) .
\end{equation}
Now $I_B$ is half the contribution of the one-loop ladder graph, which in CMY terms is $I_B=(4
S_1)/2 = 2 S_1$.  Note that we must regulate the infrared divergence via dimensional regulation
for purposes of comparison, since that is the method of CMY.  Now the sum $I_{f5}+I_{f6}$,
with the substitutions for $I_V+I_{SE}$ and $I_B$ discussed above, is given by
\begin{equation}
I_{f5}+I_{f6} \rightarrow {1 \over {4 \epsilon^2}} + {1 \over \epsilon} \Bigl ( {{\pi^2} \over
8}-1 \Bigr ) + 5.84851(4) ,
\end{equation}
while the corresponding CMY contribution is
\begin{equation}
D_6+D_7+\delta m^{(1)} C_7 = {1 \over {4 \epsilon^2}} + {1 \over \epsilon} \Bigl ( {{\pi^2} \over
8}-1 \Bigr ) + 5.850(12)  .
\end{equation}

\section{Results and conclusion}

The NRQED couplings $e_1$ and $e_2$ are determined by requiring agreement between the
imaginary parts of the scattering amplitude as calculated by NRQED and QED.  The NRQED version
of this amplitude, given in Eq.~(\ref{nrqed2}), is
\begin{eqnarray}
M_{NRQED} &=& V_4^0 \Bigl \{ \Bigl [ 1+{\alpha \over \pi} e_1 + \Bigl ( {\alpha \over \pi}
\Bigr )^2 e_2 \Bigr ] + {\alpha \over \pi} \Bigl [ {{2 \pi} \over \lambda} \Bigl ( 1 + {\alpha
\over \pi} e_1 \Bigr ) - {{2 \Lambda} \over {3 m}} + {{7 \pi \lambda} \over {12}} \Bigr ] 
\nonumber \\
&+& \Bigl ( {\alpha \over \pi} \Bigr )^2 \Bigl [ {{\pi^2 (2 \ln 2+1)} \over {\lambda^2}} - {{4
\pi \Lambda} \over {3 \tilde{\lambda}}} + 2 \pi^2 \ln \Bigl ( {\Lambda \over {2
\tilde{\lambda}}} \Bigr ) + {{4 \pi^2} \over 3} \Bigr ] \Bigr \}.
\end{eqnarray}
The corresponding QED version is
\begin{eqnarray}
M_{QED} &=& V_4^0 \Bigl \{ 1 + {\alpha \over \pi} \Bigl [ {{2 \pi} \over \lambda} + A(\lambda)
\Bigr ] \nonumber \\
&+& \Bigl ( {\alpha \over \pi} \Bigr )^2 \Bigl [ {{\pi^2 (2 \ln 2 +1)} \over
{\lambda^2}} + {{2 \pi} \over \lambda} A(\lambda) - 2 \pi^2 \ln \lambda + B_2 \Bigr ] \Bigr \}.
\end{eqnarray}
Matching at the one-loop (order $\alpha$) level yields
\begin{equation}
e_1 = {{2 \Lambda} \over {3 m}} + A(\lambda) - {{7 \pi \lambda} \over {12}}.
\end{equation}
To order $\alpha \Gamma_0$, this value of $e_1$ used in Eq.~(\ref{BSNRQED}) cancels the
ultraviolet divergent term and reproduces the Harris-Brown result \cite{Harris-Brown} after
taking the limit $\lambda \rightarrow 0$. Matching at the two-loop (order $\alpha^2$) level
gives
\begin{equation}
e_2 = -2 \pi^2 \ln \Bigl ( {\Lambda \over {2m}} \Bigr ) - {{\pi^2} \over 6} + B_2.
\end{equation}
Care is required with the photon mass $\lambda$ in this evaluation.  While it is ultimately
taken to vanish, we keep it in $e_1$ because that factor enters as a factor 
of $1/\lambda$ terms.  However, while $e_2$ also has terms proportional to the photon mass,
they can be dropped because we do not need the next renormalization constant $e_3$, which
enters in higher order.  We note that the uncalculated $O(\lambda)$ part of $A(\lambda)$
cancels from the calculation.

With $e_1$ and $e_2$ now determined from the matching calculation, we return to the bound state
NRQED calculation and use these values in Eq.~(\ref{BSNRQED}). Taking the limit $\lambda
\rightarrow 0$ then gives our final result for the two photon decay rate through two-loop
order:
\begin{equation}
\Gamma_{\rm NRQED} = \Bigl \{ 1 + {\alpha \over \pi} \Bigl ( {{\pi^2} \over 4} - 5 \Bigr ) +
\Bigl ( {\alpha \over \pi} \Bigr )^2 \bigl [ -2 \pi^2 \ln \alpha + B_{2 \gamma} \bigr ] \Bigr
\} \Gamma_0,
\end{equation}
where
\begin{equation}
B_{2 \gamma} = B_2 + {65 \pi^2 \over 24} + 2 \pi^2 \ln 2 = 5.1243(33).
\end{equation}
As  noted in the introduction, this is consistent with the CMY result.  Inclusion of all known
contributions leads to
\begin{equation}
\Gamma_{p-{\rm Ps}} = \Bigl \{ 1 + A {\alpha \over \pi} -2 \alpha^2 \ln \alpha
+B \Bigl ( {\alpha \over \pi} \Bigr )^2
- {{3 \alpha^3} \over {2 \pi}} \ln^2 \alpha + C {{\alpha^3} \over \pi} \ln \alpha
+D \Bigl ( {\alpha \over \pi} \Bigr )^3 \Bigr \} \Gamma_0,
\end{equation}
where $A=\pi^2/4-5$, 
\begin{equation}
B = B_{2 \gamma} + B_{4 \gamma} = 5.3986(33),
\end{equation}
$C=7.9189$, and $D$ is uncalculated.  Our final numerical result is
\begin{equation}
\Gamma_{p-{\rm Ps}} = 7989.6178(2) \mu s^{-1}.
\end{equation}
The uncalculated $D$ term makes a contribution of $0.00010 D \mu s^{-1}$.  Numerical
contributions of the various terms in $\Gamma_{p-{\rm Ps}}$ are detailed in Table~VIII.  While
the experimental precision is relatively high for a decay measurement, it is far too low to be
sensitive to the precise value of $B$, only disfavoring a very large value on the order of 50
or more. The very short lifetime of the state makes it unlikely that significant improvements
in the experimental precision can be achieved in the near future.

We consider one of the most important aspects of the present calculation to be its impact
on the decay rate of $o$-Ps. We have applied exactly the same methods 
to both decay rates. In the $o$-Ps case, while no independent check of our calculation
was then, or is now, available, we did use the same implementation of NRQED to calculate
a contribution to the ground state hyperfine splitting of positronium that had been
determined using Coulomb gauge Bethe-Salpeter methods \cite{Adkins} and additionally an
independent NRQED approach \cite{Hoang}, finding good agreement. This indirect  confirmation
of the validity of our methods has now been further buttressed by the fact that we have
agreement for the decay rate of $p$-Ps with CMY, who, as we have emphasized above, use an
entirely different implementation of NRQED. Therefore, the confrontation of QED with
experiment in this system, summarized as
\begin{eqnarray}
\Gamma_{o-{\rm Ps}}^{th}  [13]     & = &   7.039979(11) \mu s^{-1}     \nonumber \\
\Gamma_{o-{\rm Ps}}^{exp} [35]     & = &   7.0514(14) \mu s^{-1}       \nonumber \\
\Gamma_{o-{\rm Ps}}^{exp} [36]     & = &   7.0482(16) \mu s^{-1}       \nonumber \\
\Gamma_{o-{\rm Ps}}^{exp} [37]     & = &   7.0398(29) \mu s^{-1}          
\end{eqnarray}
remains in an unresolved state, and strongly indicates the need for further experimental work.

\acknowledgments

The work of JS was partially supported by NSF grant PHY-0097641, and that
of GA by NSF grant PHY-0070819. GA acknowledges the hospitality of the Physics
department of UCLA.  The  constant encouragement of the Michigan experimental
group is gratefully acknowledged.

\newpage

\begin{table}
\caption{Unrenormalized contributions to the parapositronium decay rate from
class a.  Class a consists of two-loop vertex corrections.}
\label{class_a_results}
\begin{tabular}{cccc}
\hline
diagram & ${1 \over {\epsilon^2}}$ & ${1 \over \epsilon}$ & $1$ \\
\hline
   a1 & ${1 \over 8}$ & ${{\pi^2} \over {16}} - \ln2 + {3 \over {16}}$ &
$-$0.3237469(14) \\
   a2 & $-{1 \over 8}$  & $-{{\pi^2} \over {16}} + \ln2 - {3 \over {16}}$ &
$-$0.3245758(11) \\
   a3 & ${1 \over 8}$ & ${{\pi^2} \over {16}} - \ln2 + {5 \over {16}}$ &
1.4769869(20) \\
   a4 & $0$ & $-{1 \over 2}$ & 3.492237(3) \\
   a5 & $-{1 \over 8}$ & $-{{\pi^2} \over {16}} + \ln2 - {3 \over {16}}$ &
$-$1.330000(3) \\
   a6 & ${1 \over 8}$ & ${{\pi^2} \over {16}} - \ln2 + {3 \over {16}}$ &
$-$0.1067757(12) \\
\hline
total & ${1 \over 8}$ & ${{\pi^2} \over {16}} - \ln2 - {3 \over {16}}$ &
2.8841255(52) \\
\hline
\end{tabular}
\end{table}

\begin{table}
\caption{Unrenormalized contributions to the parapositronium decay rate from
class b.  Class b consists of two-loop self-energy corrections.}
\label{class_b_results}
\begin{tabular}{cccc}
diagram & ${1 \over {\epsilon^2}}$ & ${1 \over \epsilon}$ & $1$ \\
\hline
   b1 & $-{1 \over 8}$ & $2 \ln2 - {7 \over {16}}$ & $-$1.4022755(10) \\
   b2 & ${1 \over {16}}$ & $- \ln2 + {9 \over {32}}$ & $-$0.2704283(10) \\
\hline
total & $-{1 \over {16}}$ & $\ln2 - {5 \over {32}}$ & $-$1.6727038(15) \\
\hline
\end{tabular}
\end{table}

\begin{table}
\caption{Unrenormalized contributions to the parapositronium decay rate from
class c.  Class c consists of products of two one-loop vertex or self-energy
corrections.}
\label{class_c_results}
\begin{tabular}{cccccccccccc}
\hline
diagram &
${1 \over {\epsilon^2}}$ & ${{\pi^2} \over \epsilon}$ &
${{\ln2} \over \epsilon}$ &
${1 \over \epsilon}$ &
$\pi^4$ & $\zeta(3)$ & $\pi^2 \ln2$ & $\pi^2$ & $\ln^2 2$ & $\ln2$ & 1 \\
\hline
   c1 & $-{1 \over 4}$ & $-{1 \over {16}}$ & $3$ & $-{1 \over 2}$ & $0$ & $-{7
\over 8}$ & ${1 \over 2}$ & $-{1 \over {24}}$ & $-7$ & $6$ & $-1$ \\
   c2 & ${1 \over 8}$ & ${1 \over {16}}$ & $-1$ & $0$ & ${1 \over {128}}$ & ${7
\over 8}$ & $-{1 \over 4}$ & ${1 \over {16}}$ & $2$ & $-2$ & $0$ \\
   c3 & ${1 \over 8}$ & $0$ & $-2$ & ${1 \over 2}$ & $0$ & $0$ & $0$ & 
$-{7 \over
{48}}$ & $6$ & $-5$ & ${5 \over 4}$ \\
\hline
total & $0$ & $0$ & $0$ & $0$ & ${1 \over {128}}$ & $0$ & ${1 \over 4}$ & $-{1
\over 8}$ & $1$ & $-1$ & ${1 \over 4}$ \\
\hline
\end{tabular}
\end{table}

\begin{table}
\caption{Contributions to the parapositronium decay rate from class d.  Class d
consists of ultraviolet and infrared finite two-loop corrections.}
\label{class_d_results}
\begin{tabular}{cd}
\hline
diagram & 1 \\
\hline
   d1 & $-$0.80394(3) \\
   d2 & $-$0.07389(6) \\
\hline
total & $-$0.87783(7) \\
\hline
\end{tabular}
\end{table}

\begin{table}
\caption{Contributions to the parapositronium decay rate from class f.  Class f
contains the ladder graphs.}
\label{class_f_results}
\begin{tabular}{cccc}
\hline
diagram & $I_B(\lambda)$ & $\pi^2 \ln\lambda$ & $1$ \\
\hline
   f1-f1a0 & $I_{LS}(\lambda)$ & $-3$ & $-$55.2277(15) \\
   f2-f2a0 & $0$ & $1$ & 30.8346(7) \\
   f3 & 0 & 0 & $-$10.05855(11) \\
   f4 & 0 & 0 & 3.9965673(13) \\
   f5 & $I_V(\lambda)$ & 0 & 1.94968(4) \\
   f6 & $I_{SE}(\lambda)$ & 0 & $-$0.655722(5) \\
\hline
total & $I_{LS}(\lambda) + I_V(\lambda) + I_{SE}(\lambda)$ & -2 & 
$-$29.1611(17) \\
\hline
\end{tabular}
\end{table}

\begin{table}
\caption{Renormalized two-loop QED contributions to the 
parapositronium decay rate by class.}
\label{two-loop_results}
\begin{tabular}{cccccc}
\hline
class & ${\pi^2 \over \lambda^2}$ & ${\pi \over \lambda}$ &
$\ln^2 \lambda$ & $\ln \lambda$ & $1$ \\
\hline
a  & 0         &  0 &-2 & $-I_V(\lambda)$    & $-$1.966447(6)   \\
b  & 0         &  0 & 1 & $-I_{SE}(\lambda)$ & $-$2.0277430(15) \\
c  & 0         &  0 & 0 & 0                  &    1.274886      \\
d  & 0         &  0 & 0 & 0                  & $-$0.87783(7)    \\
e  & 0         &  0 & 0 & 0                  &    0.4473430(6)  \\
f  & $2 \ln 2$ & $A(\lambda)$ & 1 & $I_V(\lambda) + I_{SE}(\lambda) - 
2 \pi^2$ &
$-$35.0357(32) \\
g  & 1         & $A(\lambda)$ & 0 & 0        &    1.603514      \\
h  & 0         &  0 & 0 & 0                  &    1.29392(4) \\
\hline
total&$2 \ln 2+1$&  $2A(\lambda)$ & 0 & $-2 \pi^2$ & $-$35.2881(33) \\
\hline
\end{tabular}
\end{table}

\begin{table}
\caption{Cross-check of results between this work (AMFS) and CMY.  The $1/\epsilon^2$ and
$1/\epsilon$ terms agree analytically in all cases, only the finite parts are displayed.  The
comparison for f5+f6 is subtle and is discussed in the text.}
\label{comparison}
\begin{tabular}{ccccc}
\hline
AMFS  & CMY   & AMFS   & CMY    & Difference \\
Graph & Graph & Result & Result & AMFS-CMY   \\
\hline
a1 & $D_8$                             & -0.3237469(14)  & -0.324(1)      &  0.000(1)  \\
a2 & $D_{11}+\delta m^{(1)} C_{11}$    & -0.3245758(11)  & -0.33(3)       &  0.01(3)   \\
a3 & $D_9$                             &  1.4769869(20)  &  1.475(50)     &  0.00(5)   \\
a4 & $D_{10}$                          &  3.492237(3)    &  3.488(2)      &  0.004(2)  \\
a5 & $D_{12}+\delta m^{(1)} C_{12}$    & -1.330000(3)    & -1.27(12)      & -0.06(12)  \\
a6 & $D_{13}$                          & -0.1067757(12)  & -0.12(1)       &  0.01(1)   \\
b1 & $D_{16}+\delta m^{(2)}_{CR} B_1$  & -1.4022755(10)  & -1.403(4)      &  0.001(4)  \\
b2 & $D_{17}+\delta m^{(1)} C_{17}
             +\delta m^{(2)}_{DR} B_1$ & -0.2704283(10)  & -0.271(1)      &  0.001(1)  \\
c1 & $D_{19}+\delta m^{(1)} C_{19}$    & {\rm analytic}  & {\rm analytic} &  0         \\
c2 & $D_{18}$                          & {\rm analytic}  & {\rm analytic} &  0         \\
c3 & $D_{15}+\delta m^{(1)} C_{15}
              +(\delta m^{(1)})^2 B_2$ & {\rm analytic}  & {\rm analytic} &  0         \\
d1 & $D_{14}$                          & -0.80394(3)     & -0.804(2)      &  0.000(2)  \\
d2 & $D_3$                             & -0.07389(6)     & -0.074(2)      &  0.000(2)  \\
f3+f4 & $D_2+D_5+\delta m^{(1)} C_5$   & -6.06198(11)    & -6.03(20)      & -0.03(20)  \\
f5+f6 & $D_6+D_7+\delta m^{(1)} C_7$   & 5.84851(4)      &  5.850(12)     & -0.001(12) \\
\hline
\end{tabular}
\end{table}

\begin{table}
\caption{Numerical values of contributions to the parapositronium decay rate.}
\label{num_contribs_table}
\begin{tabular}{ll}
\hline
term & contribution (in $\mu s^{-1}$) \\
\tableline
1 & $8032.5028(1)$ \\
$A {\alpha \over \pi}$ & -47.2534 \\
$-2 \alpha^2 \ln \alpha$ & 4.2092 \\
$B \bigl ( {\alpha \over \pi} \bigr )^2$ & 0.2340(1) \\
$-{{3 \alpha^3} \over {2 \pi}} \ln^2 \alpha$ & -0.0361 \\
$C {{\alpha^3} \over \pi} \ln \alpha$ & -0.0387 \\
$D \bigl ( {\alpha \over \pi} \bigr )^3$ & 0.00010 D \\
\tableline
total & 7989.6178(2) + 0.00010D \\
\hline
\end{tabular}
\end{table}

\end{document}